\documentclass{llncs}
\usepackage{url}

\usepackage{graphicx}
\usepackage{amsfonts}
\usepackage{hyperref}
\usepackage{enumitem}
\usepackage{bbding}
\usepackage{amsmath}

\begin{document}
\mainmatter              
\title{Verification of BSF Parallel Computational Model}
\titlerunning{Verification of BSF Parallel Computational Model}
\author{{Nadezhda Ezhova\Envelope}\thanks{The work was partially
supported by the Government of the Russian Federation according
to Act 211  (contract \mbox{No.~02.A03.21.0011.}) and the Ministry of education
and science of Russian Federation (government order 2.7905.2017/8.9).}}
\authorrunning{Nadezhda Ezhova \textit{et al.}} 
%
\tocauthor{Nadezhda Ezhova}
\institute{South Ural State University \\76 Lenin prospekt, Chelyabinsk, Russia, 454080\\
\email{ezhovana@susu.ru}}

\maketitle              

\begin{abstract}
The article is devoted to the verification of the BSF parallel computing model. The BSF\nobreakdash-\hspace{0pt}model is an evolution of the "master-slave" model and SPMD\nobreakdash-\hspace{0pt}model. The BSF\nobreakdash-\hspace{0pt}model is oriented to iterative algorithms that are implemented in cluster computing systems. The article briefly describes the basics of the BSF\nobreakdash-\hspace{0pt}model and its cost metrics. The structure of the BSF program is shown in the form of a UML activity diagram. The simulator of BSF-programs, implemented in C++ language using the MPI\nobreakdash-\hspace{0pt}library, is described. The results of computational experiments confirming the adequacy of the cost metrics of the BSF\nobreakdash-\hspace{0pt}model are presented.

\end{abstract}

\section{Introduction}
\begin{figure}
  \centering
  \includegraphics[scale=1]{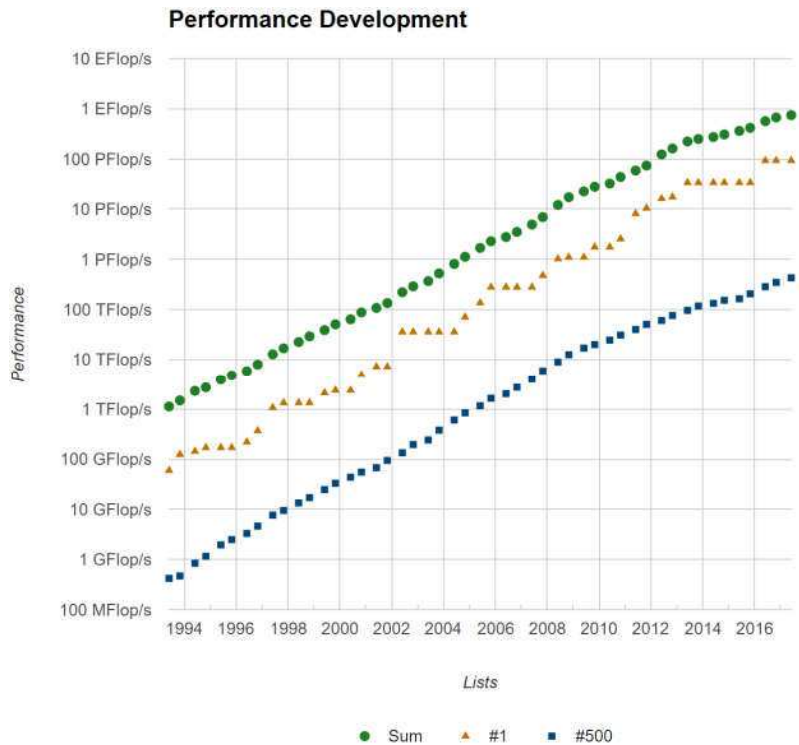}
  \caption{Performance dynamics of supercomputers in TOP500.}
  \label{figure1}
\end{figure}

The Sunway TaihuLight is the fastest supercomputer in the world according to the TOP500 List (edition of June 2017)~\cite{top500}. It uses 40,960~processors, each of which contains 260~ processing cores. The total system memory is 1.3~Petabytes; the peak performance exceeds 120~ petaflops. Analysis of the growth dynamics of the performance of supercomputers (see Fig.\,\ref{figure1}) shows that in 8\nobreakdash-\hspace{0pt}9 years the most powerful supercomputer becomes an ordinary system, and that in 5-6~years we can expect the appearance of exascale computing. The emergence of such powerful multiprocessor computing systems brings to the fore issues related to the development of frameworks (templates) that allow creating highly scalable parallel programs oriented to systems with distributed memory. In this context, the most important problem is the development of parallel computation models allowing to estimate the scalability of the algorithm at an early stage of implementing. In paper~\cite{sokol}, a new parallel computation model called BSF (Bulk-Synchronous Farm) was proposed. The BSF\nobreakdash-\hspace{0pt}model is an evolution of the "master-slave" model and focused on iterative algorithms executing on the cluster computing systems. In this paper, the issue concerning to a verification of the BSF model is discussed. The paper has the following structure. Section~2 is the description of the BSF\nobreakdash-\hspace{0pt}model. In Section~3, a cost metric of BSF\nobreakdash-\hspace{0pt}model is given. In Section~4, the simulator of BSF\nobreakdash-\hspace{0pt}programs is described and computational experiments confirming the adequacy of the cost metrics of the BSF\nobreakdash-\hspace{0pt}model are presented. In Section~5, the results are summarized and directions for further research are outlined.

\section{BSF-model}

\begin{figure}
  \centering
  \includegraphics{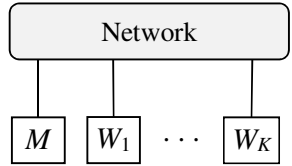}
  \caption{BSF-computer. $M$ -- master; $W_1$, \ldots, $W_K$ --– slaves.}
  \label{figure2}
\end{figure}

\emph{Parallel computation model BSF (Bulk Synchronous Farm)} was proposed in~\cite{sokol}. It extends the “Master\nobreakdash-\hspace{0pt}slave” paradigm~\cite{sahni,silva,leung} and the BSP computational model~\cite{valiant}. The BSF\nobreakdash-\hspace{0pt}model is oriented on multiprocessor systems. BSF\nobreakdash-\hspace{0pt}computer is a set of homogeneous processor nodes with private memory connected by the network that provides data transfer from one node to another. Among the processor nodes, there is one called a \emph{master} node. The remaining K nodes are called \emph{slaves}. \emph{BSF\nobreakdash-\hspace{0pt}computer} must have at least one master node and one slave node ($K \geq 1$). The architecture of the BSF\nobreakdash-\hspace{0pt}computer is shown in Fig.\,\ref{figure2}.

BSF\nobreakdash-\hspace{0pt}computer works according to the \emph{SPMD} programming model~\cite{silva,darema}. BSF\nobreakdash-\hspace{0pt}program consists of a sequence of macro-steps and global barrier synchronizations. Each macro-step is divided into sections of two types: \emph{master sections} and \emph{slave sections}. The order of such sections within the macro-step is not significant. The data processed by the particular slave is determined by the number of its node.
BSF\nobreakdash-\hspace{0pt}program includes the following sequential sections (see Fig.\,\ref{figure3}):

\begin{figure}
  \centering
  \includegraphics{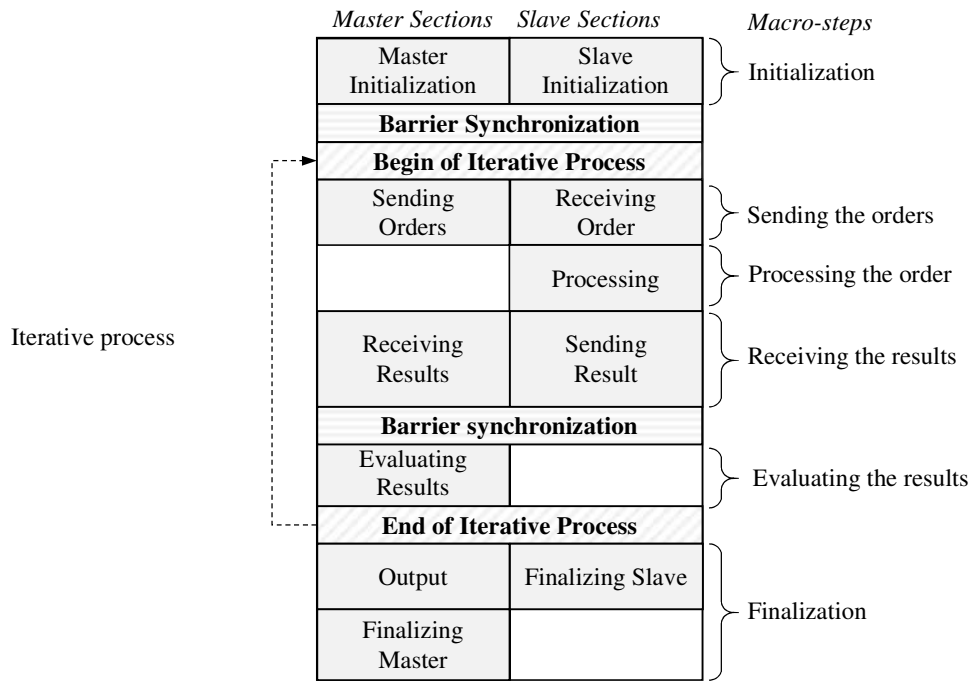}
  \caption{BSF-program structure.}
  \label{figure3}
\end{figure}

\begin{itemize}
  \item initialization;
  \item iterative process;
  \item finalization.
\end{itemize}

\emph{Initialization} is a macro\nobreakdash-\hspace{0pt}step, during which the master and slaves read or generate input data. The initialization is completed by a barrier synchronization. The \emph{iterative process} repeatedly performs its body until the exit condition checked by the master becomes true. In the \emph{finalization} macro\nobreakdash-\hspace{0pt}step, the master outputs the results and ends the program.
\emph{Body of the iterative process} includes the following macro\nobreakdash-\hspace{0pt}steps:

\begin{enumerate}
  \item [1)] sending the order (from master to slaves);
  \item [2)] processing the order (slaves);
  \item [3)] receiving the results (from slaves to master);
  \item [4)] evaluating the results (master).
\end{enumerate}

In the first macro-step, the master sends the same orders to all the slaves. Then, the slaves execute the received orders (the master is idle at that time). All the slaves execute the same program code but act on the different data with the base address depending on the slave\nobreakdash-\hspace{0pt}node number.

\begin{figure}
  \centering
  \includegraphics[scale=0.75]{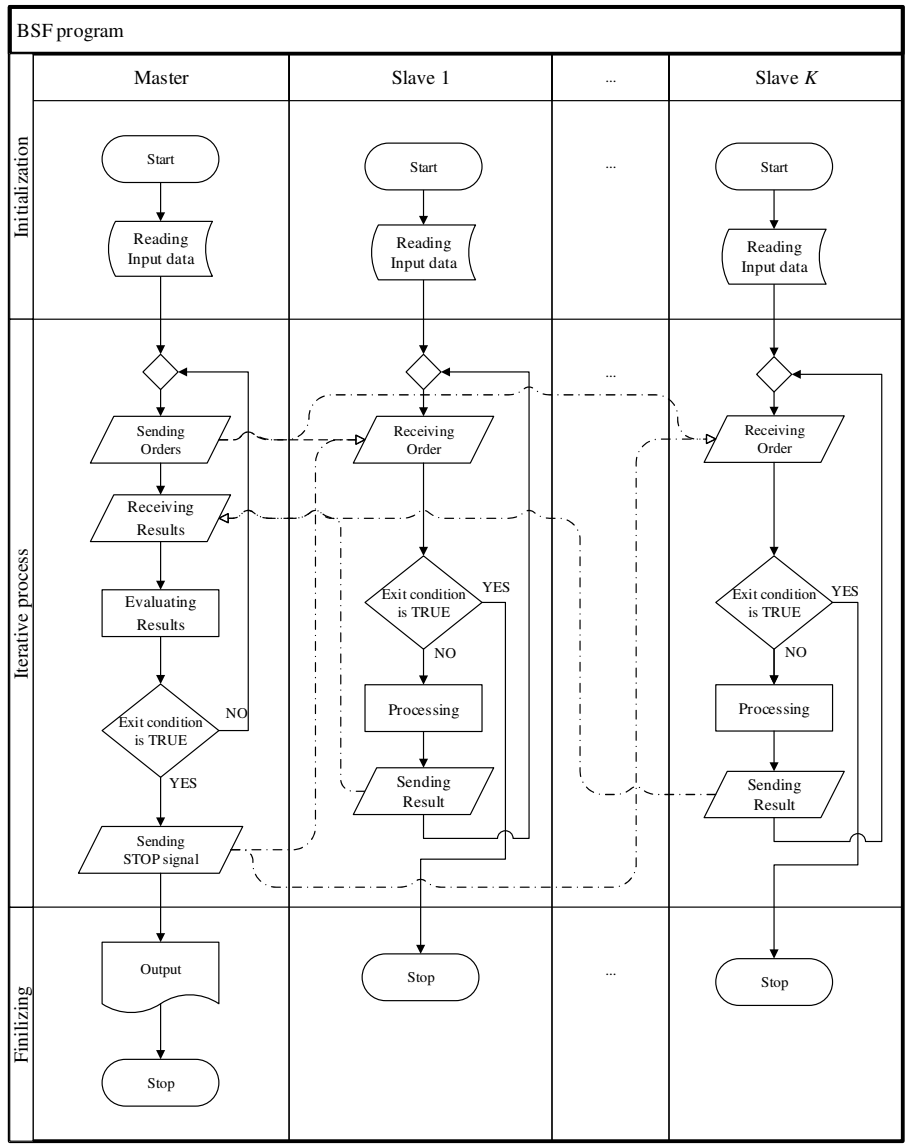}
  \caption{BSF-program structure.}
  \label{figure4}
\end{figure}

It means that all slaves spend the same time for calculating. During processing the order, there are no data transfers between nodes. The last is an important property of the BSF\nobreakdash-\hspace{0pt}model. In the third macro\nobreakdash-\hspace{0pt}step, all slaves send the results to the master. After that, global barrier synchronization is performed. During the last macro-step of iterative process, the master evaluates received results. The slaves are idle at that time. After result evaluations, the master checks the exit condition. If the exit condition is true then iterative process is finished, otherwise the iterative process is continued.
The outline of the BSF\nobreakdash-\hspace{0pt}program in form of UML activity diagram is shown in~Fig.\,\ref{figure4}.

The \emph{BSF\nobreakdash-\hspace{0pt}model} was designed for the scalable iterative numerical methods that have a high computational complexity of iteration with relatively low cost of communications. A scalable iterative method is a method that allows iteration to be splitted into subtasks that do not require data exchanges. An example of such a method can be found in~\cite{sokol_sokol}.

\section{Cost Metric of BSF\nobreakdash-\hspace{0pt}model}

The BSF\nobreakdash-\hspace{0pt}model provides an analytical estimation of the scalability of a BSF\nobreakdash-\hspace{0pt}program. The main parameters of the model are~\cite{sokol}:
\begin{itemize}
  \item [$K$:] the number of slave-nodes;
  \item [$L$:] an upper bound on the latency, or delay, incurred in communicating a message containing one byte from its source node to its target node;
  \item [${{t}_{s}}$:] time that the master-node is engaged in sending one order to one slave-node excluding the latency;
  \item [${{t}_{v}}$:] time that a slave-node is engaged in execution an order within one iteration (BSF\nobreakdash-\hspace{0pt}model assumes that this time is the same for all the slave-nodes and it is a constant within the iterative process;
  \item [${{t}_{r}}$:] total time that the master-node is engaged in receiving the results from all the slave-nodes excluding the latency;
  \item [${{t}_{p}}$:] total time that the master-node is engaged in evaluating the results received from all the slave-nodes.
\end{itemize}

Let\textsc{\char13}s denote ${{t}_{w}}=K\cdot {{t}_{v}}$ - summarized time which is spent by slave-nodes for order executions (without taking into account the effect of parallelization). Then, the upper bound of a BSF\nobreakdash-\hspace{0pt}program scalability can be estimated by the following inequality~\cite{sokol}:

\begin{equation} \label{eq1}
  K\le \sqrt{\frac{{{t}_{w}}}{2L+{{t}_{s}}}}
\end{equation}

The speedup of BSF\nobreakdash-\hspace{0pt}program can be calculated by the following equation~\cite{sokol}:

\begin{equation} \label{eq2}
    {a} = {\frac{K(2L+{{t}_{s}}+{{t}_{r}}+{{t}_{p}}+{{t}_{w}})}{{{K}^{2}}(2L+{{t}_{s}})+K({{t}_{r}}+{{t}_{p}})+{{t}_{w}}}.}
\end{equation}

One more important property of a parallel program is the parallel efficiency. The parallel efficiency of a BSF\nobreakdash-\hspace{0pt}program can be calculated by the following approximate equation~\cite{sokol}:

\begin{equation} \label{eq3}
	e\approx \frac{1}{{1+\left( {{K}^{2}}(2L+{{t}_{s}})+K({{t}_{r}}+{{t}_{p}}) \right)}/{{{t}_{w}}}\;}.
\end{equation}

\section{Verification of BSF-model}

To verify the cost metrics of the BSF\nobreakdash-\hspace{0pt}model, a simulator of BSF\nobreakdash-\hspace{0pt}programs was designed and implemented in C++ with MPI\nobreakdash-\hspace{0pt}library. The source code of the simulator is freely available on Github, at https://github.com/nadezhda-ezhova/BSF-simulator. The simulator uses SPMD\nobreakdash-\hspace{0pt}model and includes both the master and slave sections. The master section is performed if \verb"MPI_Comm_rank" returns 0. The slave section is performed if \verb"MPI_Comm_rank returns" a value greater than 0. The order is an arbitrary string of specified length transferring from the master to slave by the \verb"MPI_Isend" command. The slave reads the order by the \verb"MPI_Irecv" command. The order processing is simulated by calling the $usleep(t_v)$ function which causes the slave MPI\nobreakdash-\hspace{0pt}process to be suspended from execution until the specified number of real-time microseconds has elapsed. After processing, the master and all the slaves perform global barrier synchronization by using \verb"MPI_Barrier" command. As a result, the slave sends to the master an arbitrary string of specified length using \verb"MPI_Send" command. The master reads the results of all the slaves by \verb"MPI_Recv" command. The synchronization is performed by \verb"MPI_Waitall" command. Then, the master simulates evaluation of the results by calling the $usleep(t_p)$ function which causes the master MPI\nobreakdash-\hspace{0pt}process to be suspended from execution until the specified number of real-time microseconds has elapsed.

To verify the equation~(\ref{eq2}) determining the speedup, the parameter $v=\lg \left( {{{t}_{w}}}/{{{t}_{s}}}\; \right)$ connecting the values of ${{t}_{w}}$ and ${{t}_{s}}$ was introduced. The following values of the parameter were studied: 4, 4.5 and 6. The speedup curves plotted using the equation~(\ref{eq2}), were compared with the curves obtained as a result of numerical simulations performed via the BSF\nobreakdash-\hspace{0pt}simulator. Parameters of the experiments are given in Table~(\ref{table1}).

\begin{table}[hb!]
\caption{BSF parameters}
\label{table1}
\begin{center}
\begin{tabular}{|c|p{0.45\linewidth}|c|}
\hline
Parameter & \centering Semantics & Value (seconds) \\
\hline
${{t}_{r}}$ & \centering time that the master-node is engaged in sending one order to one slave-node excluding the latency & 0.01 \\
\hline
${{t}_{p}}$ & total time that the master-node is engaged in evaluating the results received from all the slave-nodes & 4.99 \\
\hline
${{t}_{w}}$ & total time that slave-nodes are engaged in execution of an orders & 500 \\
\hline
$L$ & an upper bound on the latency, or delay, incurred in communicating a message containing one byte from its source node to its target node & $2\cdot {{10}^{-5}}$ \\
\hline
\end{tabular}
\end{center}
\end{table}

The latency value $L$ was obtained experimentally as the transfer time of the message in 1 byte using the functions \verb"MPI_Send" and \verb"MPI_Recv". The value of ${{t}_{r}}$ corresponds to sending a message with a length of 14 MB. The values of the parameter ${{t}_{s}}$ depends on $v$ and ${{t}_{w}}$. This dependence is determined by the equation ${{t}_{s}}={{10}^{-v}}{{t}_{w}}$. The corresponding values of the parameter were obtained by varying the length of the order (see Table~(\ref{table2})).

\begin{table}[hb!]
\caption{Adjustment of parameter $v$}
\label{table2}
\begin{center}
\begin{tabular}{|c|c|c|}
\hline
$v$ & ${{t}_{s}}$ (seconds) & Order length \\
\hline
4 & 0.0206978 & 60 MB \\
\hline
4.5 & 0.0158160 & 6 MB \\
\hline
6 & 0.00048 & 200 KB \\
\hline
\end{tabular}
\end{center}
\end{table}

\begin{figure}
\begin{minipage}{0.49\linewidth}
\center{\includegraphics[scale=0.8]{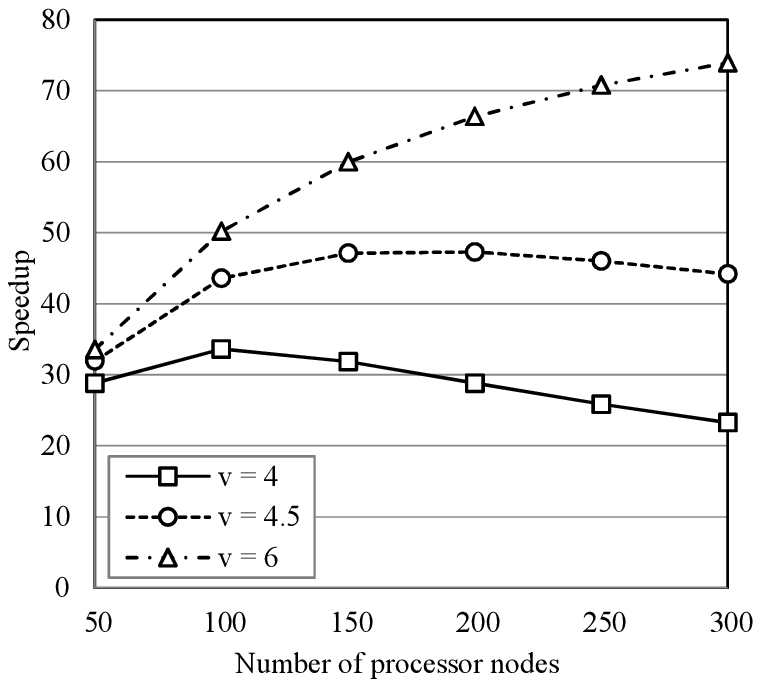}\\ a) Analytical}
\end{minipage}
\hfill
\begin{minipage}{0.49\linewidth}
\center{\includegraphics[scale=0.8]{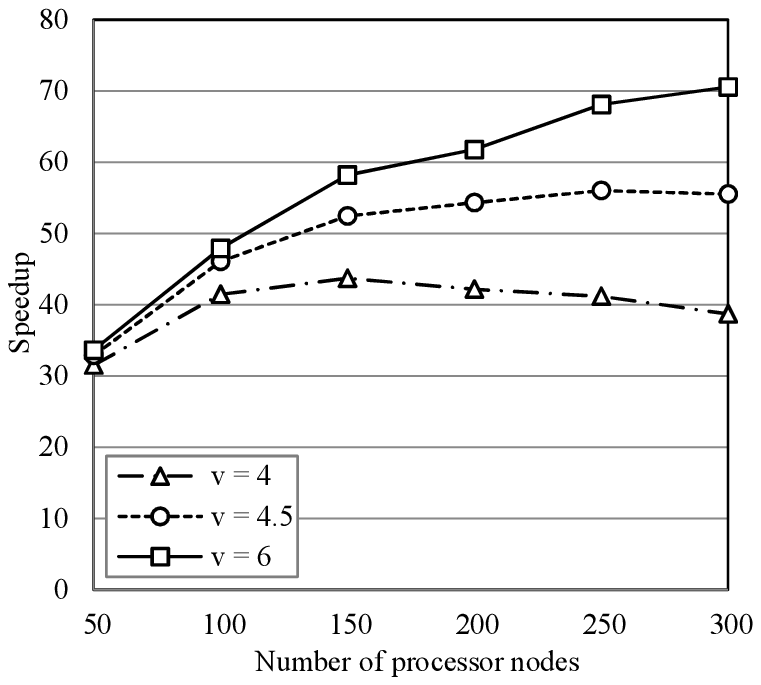}\\ b) Experimental}
\end{minipage}
\caption{Dependence of the speedup on the number of processor nodes}
\label{figure5}
\end{figure}

All the numerical experiments were conducted on the supercomputer "Tornado~SUSU"~\cite{kos_saf}. The verification results of equation~(\ref{eq2}), shown in~Fig.\,\ref{figure5}, show that the cost metrics of the BSF\nobreakdash-\hspace{0pt}model quite well predict the results produced by the BSF\nobreakdash-\hspace{0pt}simulator. In this case, the accuracy of the analytical estimates of the speedup increases with increasing of the parameter~$v$ value.

\begin{figure}
\begin{minipage}{0.49\linewidth}
\center{\includegraphics[scale=0.8]{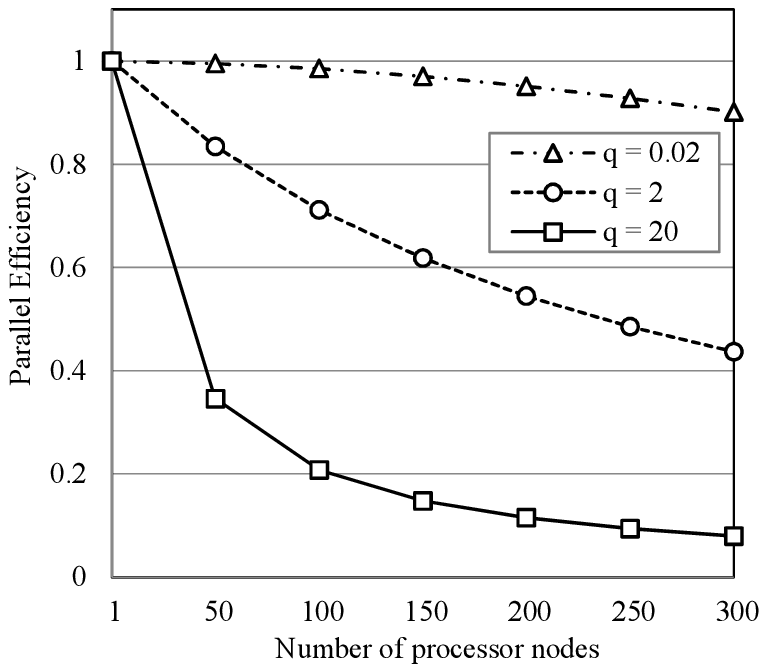}\\ a) Analytical}
\end{minipage}
\hfill
\begin{minipage}{0.49\linewidth}
\center{\includegraphics[scale=0.8]{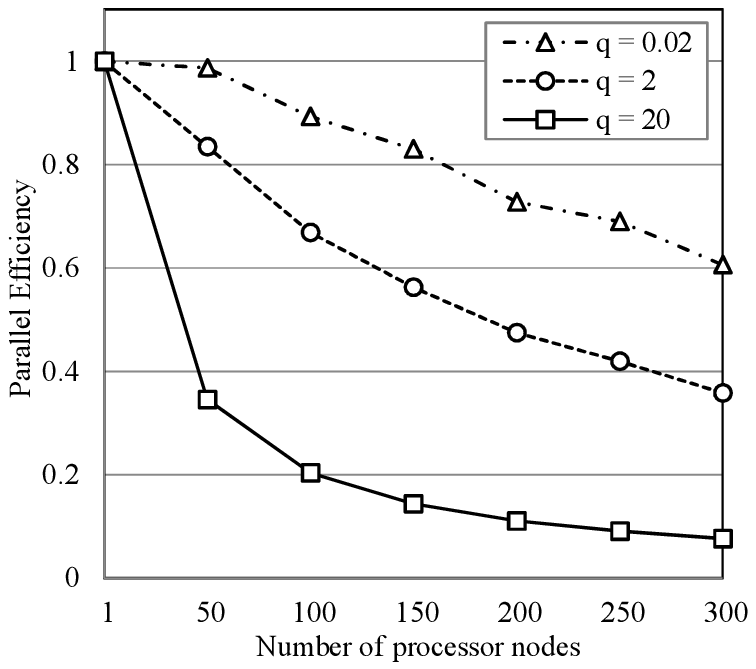}\\ b) Experimental}
\end{minipage}
\caption{Dependence of the parallel efficiency on the number of processor nodes ($v$=5)}
\label{figure6}
\end{figure}

To verify the equation~(\ref{eq3}) determining the parallel efficiency, the parameter $q={{t}_{p}}+{{t}_{r}}$ connecting the values of~${{t}_{p}}$~and~${{t}_{r}}$ was introduced. The following parameter $q$ values ​​were studied: 0.02,~2~and~20. The~parameter~${{t}_{r}}$ value was a constant equal to~0.01, and the parameter ${{t}_{p}}$ took values 0.01,~1.99~and~19.99. The parameter ${{t}_{s}}$ was also a constant equal to~0.005. This corresponds to a value of $v$ equal to~5. The verification results of equation~(\ref{eq3}), shown in~Fig.\,\ref{figure6}, show that the cost metrics of the BSF\nobreakdash-\hspace{0pt}model quite well predict the results produced by the BSF\nobreakdash-\hspace{0pt}simulator. At the same time, the accuracy of analytical estimates of parallel efficiency increases with increasing parameter~$q$ value.

\section{Conclusion}

In the paper, the issue of the adequacy of the BSF parallel computing model was studied. To verify the model, a simulator of BSF\nobreakdash-\hspace{0pt}programs in C++ language was developed using the MPI library. The emulator is implemented on the basis of the "master-slave" model. As orders, the master sends slaves messages of a certain length with arbitrary symbols. As a result, slaves are send to the master message of a certain length with arbitrary symbols. Calculations are simulated using the $usleep()$ function. Via the simulator, computational experiments were carried out for various parameters of the BSF\nobreakdash-\hspace{0pt}model. The experimental and analytical speedup curves are compared. The same comparison is made for curves of parallel efficiency. The obtained results confirm the adequacy of the BSF\nobreakdash-\hspace{0pt}model.

As future directions of research the following work is planned:
\begin{enumerate}
  \item [1)] design a skeleton for the rapid creation of BSF\nobreakdash-\hspace{0pt}applications and implement this framework in C++ using the MPI library;
  \item [2)] construct a (construct) dialog editor for fast creation of BSF\nobreakdash-\hspace{0pt}applications based on BSF-skeleton;
  \item [3)] implement some numerical methods using BSF\nobreakdash-\hspace{0pt}skeleton.
\end{enumerate}

\end{document}